\documentclass[12pt]{article}
\usepackage{lscape}
\usepackage{citesort}
\usepackage{amssymb}
\usepackage{appendix}
\usepackage{multirow}

\begin{document}

\def\qq{\langle \bar q q \rangle}
\def\uu{\langle \bar u u \rangle}
\def\dd{\langle \bar d d \rangle}
\def\sp{\langle \bar s s \rangle}
\def\GG{\langle g_s^2 G^2 \rangle}
\def\Tr{\mbox{Tr}}
\def\figt#1#2#3{
        \begin{figure}
        $\left. \right.$
        \vspace*{-2cm}
        \begin{center}
        \includegraphics[width=10cm]{#1}
        \end{center}
        \vspace*{-0.2cm}
        \caption{#3}
        \label{#2}
        \end{figure}
	}
	
\def\figb#1#2#3{
        \begin{figure}
        $\left. \right.$
        \vspace*{-1cm}
        \begin{center}
        \includegraphics[width=10cm]{#1}
        \end{center}
        \vspace*{-0.2cm}
        \caption{#3}
        \label{#2}
        \end{figure}
                }

\def\ds{\displaystyle}
\def\beq{\begin{equation}}
\def\eeq{\end{equation}}
\def\bea{\begin{eqnarray}}
\def\eea{\end{eqnarray}}
\def\beeq{\begin{eqnarray}}
\def\eeeq{\end{eqnarray}}
\def\ve{\vert}
\def\vel{\left|}
\def\ver{\right|}
\def\nnb{\nonumber}
\def\ga{\left(}
\def\dr{\right)}
\def\aga{\left\{}
\def\adr{\right\}}
\def\lla{\left<}
\def\rra{\right>}
\def\rar{\rightarrow}
\def\lrar{\leftrightarrow}  
\def\nnb{\nonumber}
\def\la{\langle}
\def\ra{\rangle}
\def\ba{\begin{array}}
\def\ea{\end{array}}
\def\tr{\mbox{Tr}}
\def\ssp{{\Sigma^{*+}}}
\def\sso{{\Sigma^{*0}}}
\def\ssm{{\Sigma^{*-}}}
\def\xis0{{\Xi^{*0}}}
\def\xism{{\Xi^{*-}}}
\def\qs{\la \bar s s \ra}
\def\qu{\la \bar u u \ra}
\def\qd{\la \bar d d \ra}
\def\qq{\la \bar q q \ra}
\def\gGgG{\la g^2 G^2 \ra}
\def\gGg{m_0^2 \la \bar q q \ra}
\def\GG{\langle g_s^2 G^2 \rangle}
\def\g5{\gamma_5 \not\!q}
\def\x{\gamma_5 \not\!x}
\def\g5{\gamma_5}
\def\sb{S_Q^{cf}}
\def\sd{S_d^{be}}
\def\su{S_u^{ad}}
\def\sbp{{S}_Q^{'cf}}
\def\sdp{{S}_d^{'be}}
\def\sup{{S}_u^{'ad}}
\def\ssp{{S}_s^{'??}}

\def\sig{\sigma_{\mu \nu} \gamma_5 p^\mu q^\nu}
\def\fo{f_0(\frac{s_0}{M^2})}
\def\ffi{f_1(\frac{s_0}{M^2})}
\def\fii{f_2(\frac{s_0}{M^2})}
\def\O{{\cal O}}
\def\sl{{\Sigma^0 \Lambda}}
\def\es{\!\!\! &=& \!\!\!}
\def\ap{\!\!\! &\approx& \!\!\!}
\def\ar{&+& \!\!\!}
\def\arrr{\!\!\!\! &+& \!\!\!}
\def\ek{&-& \!\!\!}
\def\vev{&\vert& \!\!\!}
\def\kek{\!\!\!\!&-& \!\!\!}
\def\cp{&\times& \!\!\!}
\def\se{\!\!\! &\simeq& \!\!\!}
\def\eqv{&\equiv& \!\!\!}
\def\kpm{&\pm& \!\!\!}
\def\kmp{&\mp& \!\!\!}
\def\mcdot{\!\cdot\!}
\def\erar{&\rightarrow&}


\def\simlt{\stackrel{<}{{}_\sim}}
\def\simgt{\stackrel{>}{{}_\sim}}

\def\ijpm{\stackrel{i=+,-}{{}_{j=+,-}}}

\title{
         {\Large
                 {\bf
Radiative decays of negative parity heavy baryons in QCD
                 }
         }
      }

\author{\vspace{1cm}\\
{
{\small A. K. Agamaliev$^1$\thanks{e-mail: guluoglu@mail.ru}}\,,
{\small T. M. Aliev$^2$\thanks{e-mail: taliev@metu.edu.tr}~\footnote{
permanent address:Institute of Physics, Baku, Azerbaijan}}\,\,,
{\small M. Savc{\i}$^2$\thanks{e-mail: savci@metu.edu.tr}}
} \\
{\small $^1$ Faculty of Physics, Baku State University, Az--1148,
Baku, Azerbaijan 
}\\
{\small $^2$ Physics Department, Middle East Technical University,
06531 Ankara, Turkey }}

\date{}

\begin{titlepage}
\maketitle
\thispagestyle{empty}

\begin{abstract}

The transition form factors responsible for the radiative
$\Sigma_Q \to \Lambda_Q \gamma$ and $\Xi_Q^\prime \to \Xi
\gamma$ decays of the negative parity baryons are examined within
light cone QCD sum rules. The decay widths of the
radiative transitions are calculated using the obtained results of the
form factors.

\end{abstract}

~~~PACS numbers: 11.55.Hx, 13.40.Hq, 14.20.Lq, 14.20.Mr

\end{titlepage}

\section{Introduction}

Baryons with single heavy quarks receive a lot of attention in the high
energy physics community. This is due particularly to the development of the
heavy quark effective theory and its application to this baryonic system.

Despite considerable development on the theoretical side, noticeable
progress has only recently been made on the experimental side. Lately, all
hadrons containing single heavy quark with positive parity, except
$\Omega_b^\ast$, as well as several heavy baryons possessing negative parity
have been discovered in experiments (for a review, see \cite{Rapo01}).

Baryons containing only a single heavy quark are usually described in terms
of SU(3) multiplets, which can be represented as a subgroup of the larger
SU(4) group including all baryons with zero, one two or three charmed
quarks. In this case baryon multiplet structures appear in the form
$4 \otimes 4\otimes 4 = 20 \oplus 20 \oplus 20 \oplus 4$. The symmetric 
subgroup 20 contains the decuplet with positive parity, the mixed--symmetric
subgroup 20 contains the octets of the lowest level with positive parity,
and the antisymmetric subgroup 4 contains $\Xi_c^0,~\Lambda_c,~\Xi_c^+$ and
$\Lambda$. Note that the singlet $\Lambda$ has the quantum number
$J^P={1\over 2}^-$. Similar construction takes place for the baryons
containing  single $b$ quark.

After experimental observation of the negative parity heavy baryons, the
next step is the study of their electromagnetic, weak and strong
decays. In this sense the study of electromagnetic decays plays exceptional
role, which provide us information about the internal structure of negative
parity baryons, as well as about the nonperturbative aspects of QCD.

The radiative decays between positive parity baryons in framework of the
relativistic quark model \cite{Rapo02}, in heavy baryon chiral perturbation
theory \cite{Rapo03}, in the formalism that incorporates both heavy quark
symmetry and chiral symmetry \cite{Rapo04}, in static quark model
\cite{Rapo05}, in bag model \cite{Rapo06}, in light cone QCD sum rules
incorporated with the heavy quark effective theory \cite{Rapo07}, etc.
The analysis of similar decays for the negative parity baryons has recently
started. Therefore the study of radiative decays of negative parity baryons
represent proves to be very useful in order to get information about their
properties.

In the present work we study the radiative decays $\Sigma_Q \to \Lambda_Q
\gamma$ and $\Xi_Q^\prime \to \Xi \gamma$ between negative
parity heavy baryons in framework of the light cone QCD sum rules method.
Note that the same transitions for the positive parity baryons have been
studied in the same method in \cite{Rapo08}.

The paper is organized as follows. In section 2, the light cone QCD sum
rules for the electromagnetic form factors responsible for the $\Sigma_Q \to
\Lambda_Q \gamma$ and $\Xi_Q^\prime \to \Xi \gamma$ transitions are derived.
The results numerical analysis of the sum rules for the form factors
is presented in Section 3. Using the values of these form factors at
evaluated $Q^2=0$ which corresponds to the real photon emission, the
corresponding decay widths are calculated.

\section{Sum rules for the transition form factors between negative parity
heavy baryons}

In this section we shall construct the light cone QCD sum rules for the
$\Sigma_Q \to  \Lambda_Q \gamma$ and $\Xi_Q^\prime \to \Xi \gamma$ transition
form factors between negative parity heavy baryons. For this purpose we
consider the following correlation function,
\bea
\label{eapo01}
\Pi_\mu (p,q) = - \int d^4x \int d^4y e^{i(px+qy)} \lla 0 \vel \mbox{\rm T} \{
\eta_{Q_1} (0) J_\mu^{el} (y) \bar{\eta}_{Q_2}(x) \}\ver 0 \rra ~,
\eea
where $\eta_{Q_1}$ and $\eta_{Q_2}$ are the interpolating currents of the
initial and final heavy baryons which interact simultaneously with both
positive and negative parity baryons, $j_\mu^{el} = e_q \bar{q} \gamma_\mu q +
e_Q \bar{Q} \gamma_\mu Q$ is the electromagnetic current with, and $e_q$ and $e_Q$
are the electric charges for the light and heavy quarks, respectively. 

The radiated photon can be absorbed into the electromagnetic background field
which is defined as $F_{\mu\nu} = i (\varepsilon_\mu q_\nu - \varepsilon_\nu q_\mu)
e^{iqx}$. The correlation function then can be written as,
\bea
\label{eapo02}
\Pi_\mu(p,q) \varepsilon^\mu = i \int d^4x e^{ipx} \lla 0 \vel T
\left\{\eta_{Q_1}(x) \, \bar{\eta}_{Q_2} (0) \right\} \ver 0 \rra_F~,   
\eea
where the subscript $F$ means that vacuum expectation values 
of the corresponding operators are evaluated in the presence of the background
field. Here we note that, Eq. (\ref{eapo01}) can be obtained by expanding the
correlation function in powers of $F_{\mu\nu}$, and keeping only terms linear
in $F_{\mu\nu}$ which corresponds to the single photon emission
(more about the background field method and its applications can be found in
\cite{Rapo09} and \cite{Rapo10}).

It follows from Eq. (\ref{eapo02}) that, in order to calculate the
correlation function, the expressions of the interpolating currents of heavy
baryons are needed. The general form of the interpolating currents of the
spin--1/2 heavy baryons entering into the symmetric sextet and antisymmetric
antitriplet representations are \cite{Rapo11},
\bea
\label{eapo03}
\eta_Q^{(6)} \es -{1\over \sqrt{2}} \epsilon^{abc} \Big\{ (q_1^{aT} C
Q^b) \gamma_5 q_2^c - (Q^{aT} C q_2^b) \gamma_5 q_1^c + \beta
(q_1^{aT} C \gamma_5 Q^b) q_2^c - \beta (Q^{aT} C \gamma_5 q_2^b) q_1^c
\Big\}~, \nnb \\
\eta_Q^{(\bar{3})} \es {1\over \sqrt{6}} \epsilon^{abc} \Big\{ 2 (q_1^{aT} C
q_2^b) \gamma_5 Q^c + (q_1^{aT} C Q^b) \gamma_5 q_2^c +
(Q^{aT} C q_2^b) \gamma_5 q_1^c \nnb \\
\ar 2 \beta (q_1^{aT} C \gamma_5 q_2^b) Q^c +
(q_1^{aT} C \gamma_5 Q^b) q_2^c +
(Q^{aT} C \gamma_5 q_2^b) q_1^c \Big\}~,
\eea
where $\beta$ is the arbitrary auxiliary parameter. The light quark
contents of the heavy baryons in sextet and antitriplet representations are
presented in Table 1.


\begin{table}[h]

\renewcommand{\arraystretch}{1.3}
\addtolength{\arraycolsep}{-0.5pt}
\small
$$
\begin{array}{|c|c|c|c|c|c|c|c|c|c|}
\hline \hline
 & \Sigma_{b(c)}^{+(++)} & \Sigma_{b(c)}^{0(+)} & \Sigma_{b(c)}^{-(0)} &
             \Xi_{b(c)}^{\prime -(0)}  & \Xi_{b(c)}^{\prime 0(+)}  &
\Xi_{b(c)}^{-(0)}  & \Xi_{b(c)}^{0(+)} & \Lambda_{b(c)}^{0(+)} \\
\hline \hline
q_1   & u & u & d & d & u & d & u & u \\
q_2   & u & d & d & s & s & s & s & d \\
\hline \hline
\end{array}
$$
\caption{Light quark contents of the heavy spin--1/2baryons.}
\renewcommand{\arraystretch}{1}
\addtolength{\arraycolsep}{-1.0pt}
\end{table}


According to the general strategy of QCD sum rules method, the correlation
function (\ref{eapo02}) should be evaluated in two different kinematical
domains. On the one side Eq. (\ref{eapo02}) should be dominated by the
decays $\Sigma_Q \to \Lambda_Q \gamma$ and  $\Xi_Q^\prime \to \Xi \gamma$
if the virtualities $p^2$ and $(p+q)^2$ are close to the heavy baryon masses,
i.e., $p^2 \simeq m_{final}^2$, $(p+q)^2 \simeq m_{initial}^2$. On the other
side, it can be calculated in the kinematical domain $p^2 \ll 0$ and
$(p+q)^2 \ll 0$ as an expansion in terms of the photon distribution
amplitudes with increasing twist. 

We proceed now calculating the correlation function from the hadronic part.
For this aim the complete set of hadrons carrying the same quantum numbers
as the interpolating current are inserted. At this point it is useful
noting that the interpolating current can interact both with positive and
negative parity baryons. Taking this fact into account, and calculating the
correlation function from the hadronic side, we get
\bea
\label{eapo04}
\Pi_\mu(p,q) \epsilon^\mu \es \varepsilon^\mu \sum_{\ijpm}
{\la 0 \ve \eta_{Q_2} \ve B_{2^i} (p,s) \ra \over p^2-m_{2^i}^2}
\lla  B_{2^i} (p,s)  \ve j_\mu^{el}\ve B_{1^j} \rra 
{\lla  B_{1^j} (p+q,s)  \ve \bar{\eta}_{Q_1} \ve 0 \rra \over (p+q)^2 -
m_{1^j}^2}~,
\eea
where $+(-)$ means positive(negative) parity baryon. 

The matrix element entering into Eq. (\ref{eapo04}) is defined as,
\bea
\label{eapo05}
\la 0 \ve \eta_{Q_2} \ve B_{2^i}(p,s)\ra \es  \lambda_i \Gamma^i u^i (p,q)~,\nnb \\
\la B_{2^i}(p,s) \ve j_\mu^{el} \ve B_{1^j}(p+q,s)\ra \es
\bar{u}^i \bigg[\bigg( \gamma_\mu - {\not\!{q} q_\mu \over q^2} \bigg)
f_1^\alpha - {i \sigma_{\mu\nu} q^\nu \over m_{1^j} + m_{2^i}}
f_2^\alpha \bigg] \Gamma^\alpha u^j(p+q,s)~,
\eea
where 
\bea
\Gamma^\alpha = \left\{ \begin{array}{c}
I,~\mbox{for the $+\to +$, and $-\to -$ transitions}~, \\
\gamma_5,~\mbox{for the $+\to -$, and $-\to +$ transitions}~.
\end{array} \right. \nnb
\eea
 
Using the equation of motion, the matrix element (\ref{eapo05}) can be written as,
\bea
\label{eapo06}
\la B_{2^i}(p,s) \ve \!\!\!&j_\mu^{el}&\!\!\! \ve B_{1^j}(p+q,s)\ra =
\bar{u}^i \Bigg\{\bigg( \gamma_\mu - {\not\!{q} q_\mu \over q^2} \bigg)
f_1^\alpha \nnb \\
&& + {\gamma_\mu \Big[ 2 \beta m_{1^j} + 2 m_{2^i}\Big] - 2 (p+q)_\mu
- 2 p_\mu\over 2 (m_{1^j} + m_{2^i}) } f_2^\alpha
\Bigg\} \Gamma^\alpha u^j(p+q,s)~,
\eea
where
\bea
\beta = \left\{ \begin{array}{c}
+1,~\mbox{for the $+\to +,~-\to -$ transitions}~, \\
-1,~\mbox{for the $+\to -,~-\to +$ transitions}~.
\end{array} \right. \nnb 
\eea

Imposing the conservation of the electromagnetic current, it can easily be shown
that the radiative decays  under consideration is described by the form factor
$f_2^-$. Using the condition $q\varepsilon=0$, we see from Eq. (\ref{eapo06})
that only the structure $p_\mu$ is needed for the estimation of the form
factor $f_2^-$.

Using Eqs. (\ref{eapo05}) and (\ref{eapo06}), and performing summation over
spins of the initial and final states of heavy baryons we obtain the
following expression of the correlation function from the hadronic side,
\bea
\label{eapo07}
\Pi_\mu \varepsilon^\mu \es
   -A(p\varepsilon) (\not\!{p}_2+m_{2^+}) (\not\!{p}_1+m_{1^+}) \nnb \\ 
\ar B(p\varepsilon) (\not\!{p}_2-m_{2^-}) (\not\!{p}_1-m_{1^-}) \nnb \\
\ar C(p\varepsilon) (\not\!{p}_2-m_{2^-}) (\not\!{p}_1+m_{1^+}) \nnb \\
\ar D(p\varepsilon) (\not\!{p}_2+m_{2^+}) (\not\!{p}_1-m_{1^-})+ \cdots~,
\eea
where dots represent contributions of higher states and continuum. $A$, $B$
$C$, and $D$ in Eq. (\ref{eapo07}) are given as,
\bea
\label{eapo08}
A \es {2 \lambda_{1^+} \lambda_{2^+} \over m_{1^+} + m_{2^+}}{f_2^+ \over (p^2-m_{2^+}^2) [(p+q)^2
-  m_{1^+}^2]}~, \nnb \\
B \es {2 \lambda_{1^-} \lambda_{2^-} \over m_{1^-} + m_{2^-}}{f_2^- \over (p^2-m_{2^-}^2) [(p+q)^2 
-  m_{1^-}^2]}~, \nnb \\
C \es {2 \lambda_{1^+} \lambda_{2^-} \over m_{1^+} + m_{2^-}}{f^T \over (p^2-m_{2^-}^2) [(p+q)^2 
-  m_{1^+}^2]}~, \nnb \\
D \es {2 \lambda_{1^-} \lambda_{2^+} \over m_{1^-} + m_{2^+}}{f^T \over (p^2-m_{2^+}^2) [(p+q)^2 
-  m_{1^-}^2]}~. \nnb
\eea

It follows from Eq. (\ref{eapo07}) that the correlation function contains
four different contributions coming from $+\to +$, $-\to -$,
$+\to -$, and $-\to +$ transitions. As has already been mentioned the
radiative decay between the negative parity baryons under consideration is
described only by the form factor $f_2^-$. Therefore, the unwanted
contributions coming from three transitions should be eliminated. This can
be achieved by solving the set of four equations that result from four
different Lorentz structures
$(p\varepsilon)\!\!\not\!{p}\!\!\not\!{q}$, $(p\varepsilon)\!\!\not\!{p}$,
$(p\varepsilon)\!\!\not\!{q}$, and $(p\varepsilon) I$.

In order to construct sum rules for the form factor $f_2^-$ calculation
of the correlation function from the QCD side is needed. The correlation
function given i Eq. (\ref{eapo02}) can be calculated from the QCD side
using the operator product expansion (OPE) over the twist of he nonlocal
operators. The expansion of the nonlocal operators up to twist--4 is
calculated in \cite{Rapo12}, which gets contribution from the two--particle
$\bar{q}q$, three--particle $\bar{q}Gq$, and four--particle $\bar{q}G^2q$,
$\bar{q}q\bar{q}q$ nonlocal operators. In the present work we take into
account the contributions coming only from two-- and three--particle
nonlocal operators. Indeed, taking higher Fock--space component into account
demands simultaneous calculations of the corrections with conformal spin
$j=5$ to both two-- and three--particle distribution amplitudes. The
contributions of higher conformal spin terms should be small. For this
reason, neglecting contributions of the four--particle nonlocal
operators is justified on the basis of an expansion in conformal spin (for
more details see \cite{Rapo12}). The long distance contributions are taken
into account by introducing the matrix elements of the two-- or
three--particle nonlocal operators between the vacuum and one particle
states. These matrix elements are determined with the help of the photon
distribution amplitudes (DAs). The parametrization of the aforementioned
matrix elements in terms of the photon DAs are given as,

\bea
\label{efrd10}
&&\langle \gamma(q) \vert  \bar q(x) \sigma_{\mu \nu} q(0) \vert  0
\rangle  = -i e_q \bar q q (\varepsilon_\mu q_\nu - \varepsilon_\nu
q_\mu) \int_0^1 du e^{i \bar u qx} \left(\chi \varphi_\gamma(u) +
\frac{x^2}{16} \mathbb{A}  (u) \right) \nnb \\ &&
-\frac{i}{2(qx)}  e_q \qq \left[x_\nu \left(\varepsilon_\mu - q_\mu
\frac{\varepsilon x}{qx}\right) - x_\mu \left(\varepsilon_\nu -
q_\nu \frac{\varepsilon x}{q x}\right) \right] \int_0^1 du e^{i \bar
u q x} h_\gamma(u)
\nnb \\
&&\langle \gamma(q) \vert  \bar q(x) \gamma_\mu q(0) \vert 0 \rangle
= e_q f_{3 \gamma} \left(\varepsilon_\mu - q_\mu \frac{\varepsilon
x}{q x} \right) \int_0^1 du e^{i \bar u q x} \psi^v(u)
\nnb \\
&&\langle \gamma(q) \vert \bar q(x) \gamma_\mu \gamma_5 q(0) \vert 0
\rangle  = - \frac{1}{4} e_q f_{3 \gamma} \epsilon_{\mu \nu \alpha
\beta } \varepsilon^\nu q^\alpha x^\beta \int_0^1 du e^{i \bar u q
x} \psi^a(u)
\nnb \\
&&\langle \gamma(q) | \bar q(x) g_s G_{\mu \nu} (v x) q(0) \vert 0
\rangle = -i e_q \qq \left(\varepsilon_\mu q_\nu - \varepsilon_\nu
q_\mu \right) \int {\cal D}\alpha_i e^{i (\alpha_{\bar q} + v
\alpha_g) q x} {\cal S}(\alpha_i)
\nnb \\
&&\langle \gamma(q) | \bar q(x) g_s \tilde G_{\mu \nu} i \gamma_5 (v
x) q(0) \vert 0 \rangle = -i e_q \qq \left(\varepsilon_\mu q_\nu -
\varepsilon_\nu q_\mu \right) \int {\cal D}\alpha_i e^{i
(\alpha_{\bar q} + v \alpha_g) q x} \tilde {\cal S}(\alpha_i)
\nnb \\
&&\langle \gamma(q) \vert \bar q(x) g_s \tilde G_{\mu \nu}(v x)
\gamma_\alpha \gamma_5 q(0) \vert 0 \rangle = e_q f_{3 \gamma}
q_\alpha (\varepsilon_\mu q_\nu - \varepsilon_\nu q_\mu) \int {\cal
D}\alpha_i e^{i (\alpha_{\bar q} + v \alpha_g) q x} {\cal
A}(\alpha_i)
\nnb \\
&&\langle \gamma(q) \vert \bar q(x) g_s G_{\mu \nu}(v x) i
\gamma_\alpha q(0) \vert 0 \rangle = e_q f_{3 \gamma} q_\alpha
(\varepsilon_\mu q_\nu - \varepsilon_\nu q_\mu) \int {\cal
D}\alpha_i e^{i (\alpha_{\bar q} + v \alpha_g) q x} {\cal
V}(\alpha_i) \nnb \\ && \langle \gamma(q) \vert \bar q(x)
\sigma_{\alpha \beta} g_s G_{\mu \nu}(v x) q(0) \vert 0 \rangle  =
e_q \qq \left\{
        \left[\left(\varepsilon_\mu - q_\mu \frac{\varepsilon x}{q x}\right)\left(g_{\alpha \nu} -
        \frac{1}{qx} (q_\alpha x_\nu + q_\nu x_\alpha)\right) \right. \right. q_\beta
\nnb \\ && -
         \left(\varepsilon_\mu - q_\mu \frac{\varepsilon x}{q x}\right)\left(g_{\beta \nu} -
        \frac{1}{qx} (q_\beta x_\nu + q_\nu x_\beta)\right) q_\alpha
\nnb \\ && -
         \left(\varepsilon_\nu - q_\nu \frac{\varepsilon x}{q x}\right)\left(g_{\alpha \mu} -
        \frac{1}{qx} (q_\alpha x_\mu + q_\mu x_\alpha)\right) q_\beta
\nnb \\ &&+
         \left. \left(\varepsilon_\nu - q_\nu \frac{\varepsilon x}{q.x}\right)\left( g_{\beta \mu} -
        \frac{1}{qx} (q_\beta x_\mu + q_\mu x_\beta)\right) q_\alpha \right]
   \int {\cal D}\alpha_i e^{i (\alpha_{\bar q} + v \alpha_g) qx} {\cal T}_1(\alpha_i)
\nnb \\ &&+
        \left[\left(\varepsilon_\alpha - q_\alpha \frac{\varepsilon x}{qx}\right)
        \left(g_{\mu \beta} - \frac{1}{qx}(q_\mu x_\beta + q_\beta x_\mu)\right) \right. q_\nu
\nnb \\ &&-
         \left(\varepsilon_\alpha - q_\alpha \frac{\varepsilon x}{qx}\right)
        \left(g_{\nu \beta} - \frac{1}{qx}(q_\nu x_\beta + q_\beta x_\nu)\right)  q_\mu
\nnb \\ && -
         \left(\varepsilon_\beta - q_\beta \frac{\varepsilon x}{qx}\right)
        \left(g_{\mu \alpha} - \frac{1}{qx}(q_\mu x_\alpha + q_\alpha x_\mu)\right) q_\nu
\nnb \\ &&+
         \left. \left(\varepsilon_\beta - q_\beta \frac{\varepsilon x}{qx}\right)
        \left(g_{\nu \alpha} - \frac{1}{qx}(q_\nu x_\alpha + q_\alpha x_\nu) \right) q_\mu
        \right]
    \int {\cal D} \alpha_i e^{i (\alpha_{\bar q} + v \alpha_g) qx} {\cal T}_2(\alpha_i)
\nnb \\ &&+
        \frac{1}{qx} (q_\mu x_\nu - q_\nu x_\mu)
        (\varepsilon_\alpha q_\beta - \varepsilon_\beta q_\alpha)
    \int {\cal D} \alpha_i e^{i (\alpha_{\bar q} + v \alpha_g) qx} {\cal T}_3(\alpha_i)
\nnb \\ &&+
        \left. \frac{1}{qx} (q_\alpha x_\beta - q_\beta x_\alpha)
        (\varepsilon_\mu q_\nu - \varepsilon_\nu q_\mu)
    \int {\cal D} \alpha_i e^{i (\alpha_{\bar q} + v \alpha_g) qx} {\cal T}_4(\alpha_i)
                        \right\}~,
\eea
where $\varphi_\gamma(u)$ is the leading twist-2, $\psi^v(u)$,
$\psi^a(u)$, ${\cal A}$ and ${\cal V}$ are the twist-3, and
$h_\gamma(u)$, $\mathbb{A}$, ${\cal T}_i$ ($i=1,~2,~3,~4$) are the
twist-4 photon DAs, and $\chi$ is the magnetic susceptibility.
The measure ${\cal D} \alpha_i$ is defined as
\bea
\label{nolabel05}
\int {\cal D} \alpha_i = \int_0^1 d \alpha_{\bar q} \int_0^1 d
\alpha_q \int_0^1 d \alpha_g \delta(1-\alpha_{\bar
q}-\alpha_q-\alpha_g)~.\nnb
\eea

Calculating the correlation function from the QCD side, and separating the
coefficients of the Lorentz structures
$(p\varepsilon)\!\!\not\!{p}\!\!\not\!{q}$, $(p\varepsilon)\!\!\not\!{p}$,
$(p\varepsilon)\!\!\not\!{q}$, and $(p\varepsilon) I$ from both hadronic and
QCD side, we get the following four equations for determination of the form
factor $f_2^-$,
\bea
\label{eapo09}
-A+B+C+D \es \Pi_1^{th}~,\nnb \\
- m_{2^+} (A - D) - m_{2^-} ( B + C ) \es \Pi_2^{th}~,\nnb \\
-(m_{1^+}+m_{2^+}) A - (m_{1^-}+m_{2^-}) B \nnb \\
+ (m_{1^+}-m_{2^-}) C
- (m_{1^-}-m_{2^+}) D \es \Pi_3^{th}~,\nnb \\
- m_{2^+} (m_{1^+} + m_{2^+}) A + m_{2^-} (m_{1^-} + m_{2^-}) B \nnb \\
- m_{2^-} (m_{1^+} - m_{2^-}) C
- m_{2^+} (m_{1^-} - m_{2^+}) D \es \Pi_4^{th}~.
\eea

Solving Eq. (\ref{eapo09}) for the form factor $f_2^-$, and performing
the Borel transformation over the variables $-p^2$ and $-(p+q)^2$ using 
\bea
{\cal B}\Bigg\{ {1\over (p_1^2-m_1^2) (p_2 - m_2^2)} \Bigg\}
\to e^{-m_1^2/M_1^2 - m_2^2/M_2^2}~, \nnb
\eea
in order to suppress higher states and continuum contributions,
we finally get the following sum rules for the form factor $f_2^-$
at the point $-q^2=Q^2=0$,
\bea
\label{eapo10}
f_2^-(Q^2=0) \es {m_{1^-} + m_{2^-}\over 2 \lambda_{1^-} \lambda_{2^-}}
e^{-m_1^2/M_1^2 - m_2^2/M_2^2} \Bigg\{  {1\over (m_{1^-} + m_{1^+})
(m_{2^-} + m_{2^+})}
\Big[(m_{1^+} - m_{2^-}) m_{2^+} \Pi_1^B \nnb \\
\ek m_{2^+} \Pi_2^B - (m_{1^+} - m_{2^-})\Pi_3^B -  
  \Pi_4^B\Big] \Bigg\}
+ \int ds_1 ds_2 \rho^h (s_1,s_2) e^{-s_1/M_1^2 -s_2/M_2^2}~.
\eea

The last term in Eq. (\ref{eapo10}) represents the contributions of the
higher states, as well as continuum. This contribution is usually estimated
by using the quark--hadron duality ansatz, which states that above some
threshold in the $(s_1,s_2)$ plane the hadronic spectral density $\rho^h
(s_1,s_2)$ is equal to the spectral density calculated from the QCD side.

The explicit expressions of $\Pi_i^B$ for the $\Sigma_Q \to \Lambda_Q \gamma$
and $\Xi_Q^\prime \to \Xi_Q \gamma$ transitions  are presented in Appendix.

Few words about the continuum subtraction procedure are in order. This
procedure is explained in detail in \cite{Rapo13}, where use has been made of
the quark--hadron duality. In the case $M_1^2=M_2^2=2 M^2$, and $u_0=1/2$
the subtraction can be carried out with the help of the formula,
\bea
M^{2n} e^{-m^2/M^2} \to {1\over \Gamma(n)} \int_{m^2}^{s_0} ds\, e^{-s/M^2}
(s-m^2)^{n-1}~,~~(n\ge 1)~,\nnb
\eea
from which for the leading twist terms we have,
\bea
M^2 e^{-m^2/M^2} \to M^2\left(e^{-m^2/M^2}-e^{-s_0/M^2}\right)~,\nnb
\eea
where $m$ is the heavy quark mass.

In the present work the subtraction procedure is not performed for the
higher order twist terms which are proportional to the zeroth or negative
powers of $M^2$, since non of these contributions are small (see for example
\cite{Rapo13}). It should be emphasized here that, in principle, single
and double dispersion integrals coming from subtraction procedure can
appear. But such terms disappear after the double Borel transformations.

The residues of the negative parity are calculated in \cite{Rapo14}.
In the transitions under consideration, the masses of the initial and final state
heavy baryons are very close to each other, and hence we set $M_1^2=M_2^2=2
M^2$.

After carrying out the numerical analysis for the form factor $f_2^-$ the
decay widths of the transitions under consideration can easily be
calculated, whose expression is given as,
\bea
\label{eapo11}
\Gamma(B_{1^-} \to B_{2^-} \gamma) = {4 \alpha \ve \vec{q} \ve^3 \over
m_{1^-}+ m_{2^-} } \ve f_2^-(0) \ve^2,
\eea
where $\alpha$ is the fine structure constant, and
\bea
\ve \vec{q} \ve = {m_{1^-}^2 - m_{2^-}^2 \over 2 m_{1^-} }~,\nnb
\eea
is the magnitude of the photon momentum. 

\section{Numerical analysis}

Present section is devoted to the numerical analysis of the sum rules
for the form factor $f_2^-(0)$. The main nonperturbative input parameters in 
the sum rules are the photon DAs, whose expressions are given in
\cite{Rapo09}. In addition to the photon DAs the sum rules contain other
input parameters such as, the quark condensate $\qq$, vacuum expectation
value of the dimension--5 operator $\gGg$, magnetic susceptibility $\chi$ of
the quark fields. In our analysis we shall use the following values these
parameters: $\uu\ve_{1~geV} = \dd\ve_{1~geV} = -(0.243)^3~GeV^3$,
$\sp\ve_{1~geV} = 0.8 \uu\ve_{1~geV}$, $m_0^2=(0.8\pm0.2)~GeV^2$ (the value
of $m_0^2$ is determined from the analysis of the two point sum rules for
baryons, as well as from the $B,B^\ast$ system)
\cite{Rapo15,Rapo16,Rapo17},
$f_{3\gamma}=-0.0039~GeV^2$ \cite{Rapo09}. The magnetic susceptibility is
calculated in numerous works \cite{Rapo18,Rapo19,Rapo20}, and in our
numerical calculations we use $\chi(1~GeV)=-0.2.85~GeV^{-2}$.

Having determined the input parameters we can now perform the numerical
analysis of the sum rule for the form factor $f_2^-(0)$. The sum rule
includes three auxiliary parameters, namely, the continuum threshold $s_0$,
Borel mass parameter $M^2$, and the arbitrary auxiliary parameter $\beta$ in
the expression of the interpolating currents. Obviously, the measurable
quantity $f_2^-(0)$ should be independent of these parameters. Therefore
we need to find such regions for all these parameters where the form factor
$f_2^-(0)$ is practically independent of them. This program can be
implemented in the following way. As far as the continuum threshold is
concerned, analysis of various sum rules shows that the difference
$\sqrt{s_0}-m$, where $m$ is the ground state mass, varies in the region
$0.3~GeV\le \sqrt{s_0}-m\le 0.8~GeV$, and in our calculations we use the
average value $\sqrt{s_0}-m = 0.5~GeV$. In determination of the domain for
the Borel mass parameter $M^2$, we demand that the following two conditions
must be satisfied: a) The upper bound is obtained by requiring that the
higher states and continuum contributions constitute at most 40\% of the
contributions coming from the perturbative parts. The lower bound is
determined by imposing the requirement that the higher twist contributions
are less than the leading twist contributions. The analysis of the sum rules
for the negative parity heavy baryons which is studied in \cite{Rapo14},
leads to the following working regions for $M^2$,
\bea
&&2.5~GeV^2 \le M^2 \le 4.0~GeV^2,~\mbox{for~
$\Sigma_c,~\Xi_c^\prime,~\Lambda_c,~\Xi_c$}~, \nnb \\
&&4.5~GeV^2 \le M^2 \le 7.0~GeV^2,~\mbox{for~
$\Sigma_b,~\Xi_b^\prime,~\Lambda_b,~\Xi_b$}~. \nnb
\eea

Finally, the domain for the auxiliary parameter $\beta$ is determined by
requesting that  the form factor $f_2^-(0)$ shows good stability
with respect to its variation in this region. As an example, in Figs. (1)
and (2) we present the dependence of the form factor $f_2^-(0)$ for
the $\Sigma_b \to \Lambda_b$ transition, on $\cos\theta$, where $\beta
=\tan\theta$, at $s_0=42~GeV^2$
and $s_0=44~GeV^2$, and at several fixed values of $M^2$. We observe from
these figures that, in the region $-1.0 \le \cos\theta \le -0.7$, the form
factor $f_2^-(0)$ seems to be practically independent to the variation in 
$\cos\theta$, and also it is insensitive to the different choices of the
values of $s_0$ and $M^2$. Performing similar analysis for all remaining
transition channels, we get the following values for the form factor
$f_2^-(0)$ which are summarized in the following table:

\bea
\label{emrt18}
f_2^-(0) = \left\{ \begin{array}{rl}
(1.2\pm 0.3)&\mbox{for }\Sigma_c^+ \to \Lambda_c^+ \gamma~, \\
(0.8\pm 0.2)&\mbox{for }\Xi_c^{\prime +} \to \Xi_c^+ \gamma~,\\
(-0.030\pm 0.008)&\mbox{for }\Xi_c^{\prime 0} \to \Xi_c^0 \gamma~,\\
(1.4 \pm 0.2)&\mbox{for }\Sigma_b^0 \to \Lambda_b^0 \gamma~,\\
(1.2\pm 0.2)&\mbox{for }\Xi_b^{\prime 0} \to \Xi_b^0 \gamma~,\\
(-0.18\pm 0.03)&\mbox{for }\Xi_b^{\prime -} \to \Xi_b^- \gamma~.
\end{array} \right. \nnb
\eea
The uncertainties coming from the errors of input parameters are taken into
account quadratically.

Using these values of the form factors $f_2^-(0)$, and Eq.
(\ref{eapo11}), we estimate the decay widths of the decays under
consideration whose values are given as,
\bea
\Gamma_{\Sigma_c^+ \to \Lambda_c^+ \gamma}   \es 1.2   \times \left(1.0  \pm 0.5\right)~keV~,\nnb \\
\Gamma_{\Xi_c^{\prime +} \to \Xi_c^+ \gamma} \es 0.7   \times \left(1.0  \pm 0.5\right)~keV~,\nnb \\
\Gamma_{\Xi_c^{\prime 0} \to \Xi_c^0 \gamma} \es 0.002 \times \left(1.00 \pm 0.5\right)~keV~,\nnb \\
\Gamma_{\Sigma_b^0 \to \Lambda_b^0 \gamma}   \es 1.4   \times \left(1.0  \pm 0.3\right)~keV~,\nnb \\  
\Gamma_{\Xi_b^{\prime 0} \to \Xi_b^0 \gamma} \es 0.6   \times \left(1.0  \pm 0.3\right)~keV~,\nnb \\
\Gamma_{\Xi_b^{\prime -} \to \Xi_b^- \gamma} \es 0.018  \times \left(1.0  \pm 0.3\right)~keV~,\nnb
\eea

The predictions presented for the transition form factors, as well as decay widths
constitute the main objective of the present work.
It follows from these results that the decay widths of $\Sigma_c^+ \to
\Lambda_c^+ \gamma$, $\Sigma_b^0 \to \Lambda_b^0 \gamma$,
$\Xi_c^{\prime +} \to \Xi_c^+ \gamma$,and $\Xi_b^{\prime 0}
\to \Xi_b^0 \gamma$ transitions are quite large and can be measured in the
near future; while the widths of 
$\Xi_c^{\prime 0} \to \Xi_c^0 \gamma$, and $\Xi_b^{\prime -} \to \Xi_b^-
\gamma$ transitions are very small. 

In conclusion, we employ light cone QCD sum rules in calculating the form
factor $f_2^-(0)$ for the magnetic--dipole transition $M1$ between the
negative parity heavy spin--1/2 baryons. Using the values of the form
factors $f_2^-(0)$ for the transitions under consideration, we also
estimate their decay widths. Our results predict that $\Sigma_c^+ \to
\Lambda_c^+ \gamma$, $\Sigma_b^0 \to \Lambda_b^0 \gamma$, $\Xi_b^{\prime 0}
\to \Xi_b^0 \gamma$, and $\Xi_c^{\prime +} \to \Xi_c^+ \gamma$ decays
have large widths and can be measured in future
experiments.

\newpage

\section*{Appendix}
\setcounter{equation}{0}
\setcounter{section}{0}

{\bf 1) Coefficient of the $(p\varepsilon)\!\!\not\!{p}\!\!\not\!{q}$ structure}
\bea
&& \Pi_1^B = \nnb \\
%
%
%
\ek {e^{-m_b^2/M^2} \over 6912 \sqrt{3} M^{10}}
(1 - \beta) f_{3\gamma} \GG m_0^2 m_b^4 (e_u \sp - e_s \uu) 
  \Big[2 (3 + \beta) \widetilde{j}_1(\psi^v) + \beta \psi^a(u_0)\Big] \nnb \\
%
%
\ar {e^{-m_b^2/M^2} \over 3456 \sqrt{3} M^8}
(1 - \beta) f_{3\gamma} \GG m_0^2 m_b^2 (e_u \sp - e_s \uu) 
   \Big[2 (3 + \beta) \widetilde{j}_1(\psi^v) + \beta \psi^a(u_0)\Big] \nnb \\
%
%
\ar {e^{-m_b^2/M^2} \over 13824 \sqrt{3} \pi^2 M^6}
(1 - \beta) \GG m_b^2 (e_u \sp - e_s \uu) 
   \Big\{3 \beta m_0^2 \nnb \\
\ar 8 f_{3\gamma} \pi^2 \Big[2 (3 + \beta)  \widetilde{j}_1(\psi^v) +
      \beta \psi^a(u_0)\Big]\Big\} \nnb \\
%
%
\ar {e^{-m_b^2/M^2} \over 96 \sqrt{3} M^4}
(1 - \beta) f_{3\gamma} m_0^2 m_b^2 (e_u \sp - e_s \uu) 
   \Big[2 (3 + \beta)  \widetilde{j}_1(\psi^v) + \beta \psi^a(u_0)\Big] \nnb \\
%
%
\ek {e^{-m_b^2/M^2} \over 4608 \sqrt{3} \pi^2 M^2}
(1 - \beta) \GG (e_s \sp - e_u \uu) \Big\{(-1 + \beta) \mathbb{A} (u_0) - 
    3 (1 + \beta) i_2({\cal S},1) \nnb \\
\ek 3 i_2(\widetilde{\cal S},1) -
    2 i_2({\cal T}_1,1) - 3 i_2({\cal T}_2,1) + 
    2 i_2({\cal T}_3,1) + 3 i_2({\cal T}_4,1) + 6 i_2({\cal S},v) + 2 i_2(\widetilde{\cal S},v) \nnb \\
\ar 4 i_2({\cal T}_2,v) - 4 i_2({\cal T}_3,v) + \beta \Big[-3 i_2(\widetilde{\cal S},1) + 2 i_2({\cal T}_1,1) - 
      3 i_2({\cal T}_2,1) - 2 i_2({\cal T}_3,1) \nnb \\
\ar 3 i_2({\cal T}_4,1) + 2 i_2({\cal S},v) + 
      6 i_2(\widetilde{\cal S},v) + 4 i_2({\cal T}_3,v) - 4 i_2({\cal T}_4,v) - 
      8 \widetilde{j}_2(h_\gamma)\Big] - 16 \widetilde{j}_2(h_\gamma)\Big\} \nnb \\
\ek {e^{-m_b^2/M^2} \over 1152 \sqrt{3} \pi^2 M^2}
 (1 - \beta) (e_u \sp - e_s \uu) \Big\{\beta \GG - 2 f_{3\gamma} m_0^2 \pi^2 
     \Big[2 (11 + 5 \beta) \widetilde{j}_1(\psi^v) \nnb \\
\ar (2 + 5 \beta) \psi^a(u_0)\Big]\Big\} \nnb \\
%
%
%
\ar {1 \over 64 \sqrt{3} \pi^2}
(1 - \beta) (e_s \sp - e_u \uu) \Big[(1 - \beta) {\cal I}_0 -  
   (5 + \beta) m_b^2 {\cal I}_1 + 2 (2 + \beta) m_b^4 {\cal I}_2\Big] 
  \Big[i_2({\cal S},1) - i_2({\cal T}_4,1)\Big] \nnb \\
%
%
\ek {1 \over 64 \sqrt{3} \pi^2}
    (1 - \beta) (e_s \sp - e_u \uu) \Big[(1 - \beta) {\cal I}_0 +
    (1 + 5 \beta) m_b^2 {\cal I}_1 - 2 (1 + 2 \beta) m_b^4 {\cal I}_2\Big]
   \Big[i_2(\widetilde{\cal S},1) + i_2({\cal T}_2,1)\Big] \nnb \\
%
%
\ek {1 \over 32 \sqrt{3} \pi^2}
(1 - \beta) (e_s \sp - e_u \uu) \Big\{(1 - \beta) m_b^2 
({\cal I}_1 - m_b^2 {\cal I}_2) 
i_2({\cal T}_1,1) - (1 - \beta) ({\cal I}_0 - m_b^2 {\cal I}_1)
i_2({\cal T}_3,1) \nnb \\
\ek m_b^2 ({\cal I}_1 - m_b^2 {\cal I}_2) \Big[(3 + \beta) i_2({\cal S},v) +
     (1 + 3 \beta) i_2(\widetilde{\cal S},v)\Big] + 2 (1 - \beta) ({\cal I}_0 -
m_b^2 {\cal I}_1) i_2({\cal T}_3,v) \Big\} \nnb \\
%
%
\ar {1 \over 32 \sqrt{3} \pi^2}
(1 - \beta) (e_s \sp - e_u \uu)     
  \Big\{\Big[(1 - \beta) {\cal I}_0 + 2 \beta m_b^2 {\cal I}_1 - (1 + \beta)
m_b^4 {\cal I}_2\Big] i_2({\cal T}_2,v) \nnb \\
\ar \Big[(1 - \beta) {\cal I}_0 - 2 m_b^2 {\cal I}_1 +  
     (1 + \beta) m_b^4 {\cal I}_2\Big] i_2({\cal T}_4,v)\Big\} \nnb\\
%
%
\ar {3 \over 128 \sqrt{3} \pi^4}
\sqrt{3} (1 - \beta^2) (e_s - e_u) m_b^3 {\cal I}_{\ell n} \nnb \\
%
%
\ar {1 \over 384 \sqrt{3} \pi^4}
(1 - \beta) {\cal I}_{-1} \Big[3 (1 + \beta) (e_s - e_u) m_b + 
   4 (1 - \beta) \pi^2 (e_s \sp - e_u \uu) \chi \varphi_\gamma(u_0) \Big] \nnb \\
%
%
\ek {e^{-m_b^2/M^2} \over 2304 \sqrt{3} \pi^2 m_b}
(1 - \beta) \Big\{36 \beta m_0^2 m_b (e_u \sp - e_s \uu) + 
   2 (3 + \beta) f_{3\gamma} \Big[(e_s - e_u ) \GG \nnb \\
\ar 96 m_b \pi^2 (e_u\sp - 
     e_s \uu) \Big]  \widetilde{j}_1(\psi^v) + 
   e_u f_{3\gamma} \Big[(1 + \beta) \GG + 96 \beta m_b \pi^2 \sp\Big] \nnb \\
\ek e_s f_{3\gamma} \Big[(1 + \beta) \GG + 96 \beta m_b \pi^2 \uu) \Big] \psi^a(u_0)\Big\} \nnb \\
%
%
\ek {1 \over 512 \sqrt{3} \pi^4}
(1 - \beta) m_b {\cal I}_1 \Big\{(1 + \beta) (e_s-e_u) (\GG + 12 m_b^4)  
   - 32 e_b m_b \pi^2 (\sp - \uu) \nnb \\
\ar 16 m_b^2 \pi^2 \Big[2 (3 + \beta) (e_s - e_u) f_{3\gamma}  \widetilde{j}_1(\psi^v) - 
     (1 - \beta) m_b (e_s \sp - e_u \uu) \chi \varphi_\gamma(u_0) \nnb \\
\ek (1 + \beta) (e_s - e_u) f_{3\gamma} \psi^a(u_0)\Big]\Big\} \nnb \\
%
%
\ar {1 \over 1536 \sqrt{3} \pi^4 m_b}
(1 - \beta) {\cal I}_0 \Big\{(1 + \beta) (e_s-e_u) (\GG + 18 m_b^4) 
     + 48 \beta e_u m_b \pi^2 \sp \nnb \\ 
\ek 48 e_b m_b \pi^2 (\sp - \uu) - 48 \beta e_s m_b \pi^2 \uu - 
   12 \pi^2 m_b \Big[(1 - \beta) (e_s \sp - e_u \uu) \mathbb{A} (u_0) \nnb \\
\ek 4 (3 + \beta) (e_s - e_u) f_{3\gamma} m_b  \widetilde{j}_1(\psi^v) + 
     4 (e_s \sp - e_u \uu) \Big(2 (2 + \beta) \widetilde{j}_2(h_\gamma) \nnb \\
\ar (1 - \beta) m_b^2 \chi \varphi_\gamma(u_0) \Big) + 2 (1 + \beta) (e_s - e_u) 
      f_{3\gamma} m_b \psi^a(u_0)\Big]\Big\} \nnb \\
%
%
\ek {1 \over 2304 \sqrt{3} \pi^4}
(1 - \beta) m_b {\cal I}_2 \Big\{3 m_b \Big[-(1 + \beta) (e_s - e_u) m_b (\GG + 3 m_b^4) \nnb \\
\ar  2 \pi^2 (e_u \sp - e_s \uu)\Big((2 - \beta) m_0^2 + 12 \beta m_b^2\Big) -
     2 \pi^2 e_b \Big((2 - \beta) m_0^2 - 12 m_b^2\Big) (\sp - \uu) \Big] \nnb \\
\ar \pi^2 \Big[-18 (1 - \beta) m_b^3 (e_s \sp - e_u \uu) \mathbb{A} (u_0)
- 6 (3 + \beta) (e_s - e_u) f_{3\gamma} (\GG \nnb \\
\ar 12 m_b^4)  \widetilde{j}_1(\psi^v) + 
     2 m_b (e_s \sp - e_u \uu)
\Big(-72 (2 + \beta) m_b^2 \widetilde{j}_2(h_\gamma) \nnb \\
\ar (1 - \beta) (\GG + 12 m_b^4) \chi \varphi_\gamma(u_0)\Big)
+3 (1 + \beta) (e_s - e_u) f_{3\gamma} (\GG + 12 m_b^4)
\psi^a(u_0)\Big]\Big\}~. \nnb \nnb \\ \nnb \\
\eea
{\bf 2) Coefficient of the $(p\varepsilon)\!\!\not\!{p}$ structure}
\bea
&& \Pi_2^B = \nnb \\
%
%
\ek {e^{-m_b^2/M^2} \over 2304 \sqrt{3} \pi^2 M^4}
(2 - \beta - \beta^2) \GG m_0^2 m_b (e_u \sp - e_s \uu) \nnb \\
%
%
\ar { e^{-m_b^2/M^2} \over 1152 \sqrt{3} \pi^2 M^2}
(2 - \beta - \beta^2) \GG m_0^2 (e_u \sp - e_s \uu) \nnb \\
%
%
\ar{ e^{-m_b^2/M^2} \over 576 \sqrt{3} \pi^2 m_b}
(1 - \beta) (2 + \beta) \GG (e_u \sp - e_s \uu) \nnb \\
\ek {1 \over 16 \sqrt{3} \pi^2}
(1 - \beta) (2 + \beta) m_b [(e_b + e_u) \sp - (e_b + e_s) \uu] {\cal I}_0 \nnb \\
\ek {1 \over 64 \sqrt{3} \pi^2}
(1 - \beta) m_b \Big\{\Big[3 (1 + \beta) e_u m_0^2 - 8 (2 + \beta) (e_b + e_u) m_b^2\Big] 
     \sp \nnb \\
\ek \Big[3 (1 + \beta) e_s m_0^2 - 8 (2 + \beta) (e_b + e_s) m_b^2\Big]
\uu\Big\} {\cal I}_1 \nnb \\
\ek {1 \over 192 \sqrt{3} \pi^2}
(1 - \beta) m_b \Big\{\Big[(2 + \beta) e_u \GG - 
      3 \Big((3 + 2 \beta) e_b + (7 + 5 \beta) e_u\Big) m_0^2 m_b^2 \nnb \\
\ar      12 (2 + \beta) (e_b + e_u) m_b^4\Big] \sp - 
\Big[(2 + \beta) e_s \GG - 3 \Big((3 + 2 \beta) e_b + (7 + 5 \beta) e_s\Big) m_0^2 m_b^2 \nnb \\
\ar 12 (2 + \beta) (e_b + e_s) m_b^4\Big] \uu\Big\} {\cal I}_2~. \nnb \\ \nnb \\
\eea
{\bf 3) Coefficient of the $(p\varepsilon)\!\!\not\!{q}$ structure}
\bea
&& \Pi_3^B = \nnb \\
%
%
&& {e^{-m_b^2/M^2} \over 6912 \sqrt{3} M^{10}}
(1 - \beta) f_{3\gamma} \GG m_0^2 m_b^5 (e_u \sp - e_s \uu) 
     \Big[4 (2+\beta) \widetilde{j}_1(\psi^v) - 
(1 + 2 \beta) \psi^a(u_0)\Big] \nnb \\
%
%
\ek {e^{-m_b^2/M^2} \over 1152 \sqrt{3} M^8}
(1 - \beta) f_{3\gamma} \GG m_0^2 m_b^3 (e_u \sp - e_s \uu)           
     \Big[4 (2+\beta) \widetilde{j}_1(\psi^v) - 
(1 + 2 \beta) \psi^a(u_0)\Big] \nnb \\ 
%
%
\ar {e^{-m_b^2/M^2} \over 13824 \sqrt{3} \pi^2 M^6}
(1 - \beta) \GG m_b (e_u \sp - e_s \uu) \Big\{3 (1 + 2 \beta) m_0^2 m_b^2 \nnb \\ 
\ar    4 f_{3\gamma} (3 m_0^2 - 2 m_b^2) \pi^2 \Big[4 (2 + \beta)  \widetilde{j}_1(\psi^v) - 
      (1 + 2 \beta) \psi^a(u_0) \Big]\Big\} \nnb \\
%
%
\ek {e^{-m_b^2/M^2} \over 4608 \sqrt{3} \pi^2 M^4}
(1 - \beta) m_b (e_u \sp - e_s \uu) \Big\{(5 + 7 \beta) \GG m_0^2 \nnb \\
\ek 8 f_{3\gamma} (\GG - 6 m_0^2 m_b^2) \pi^2 \Big[4 (2 + \beta)  \widetilde{j}_1(\psi^v) - 
     \psi^a(u_0) - 2 \beta \psi^a(u_0)\Big]\Big\} \nnb \\
%
%
\ek {e^{-m_b^2/M^2} \over 9216 \sqrt{3} \pi^2 M^2}
(1 - \beta) \GG m_b (e_s \sp - e_u \uu) \Big\{3 (1 + \beta) \mathbb{A} (u_0) + 
   4 (1 - \beta) i_2({\cal S},1) \nnb \\
\ar 4 (1 - \beta) i_2(\widetilde{\cal S},1) + 6 i_2({\cal T}_1,1) + 
     4 i_2({\cal T}_2,1) - 6 i_2({\cal T}_3,1) - 4 i_2({\cal T}_4,1) - 4 i_2(\widetilde{\cal S},v) - 
       12 i_2({\cal T}_2,v) \nnb \\
\ar 12 i_2({\cal T}_3,v) + 
     2 \beta \Big[3 i_2({\cal T}_1,1) - 2 i_2({\cal T}_2,1) - 3 i_2({\cal T}_3,1) + 
       2 i_2({\cal T}_4,1) + 2 i_2(\widetilde{\cal S},v) -2 i_2({\cal T}_2,v) \nnb \\
\ar 6 i_2({\cal T}_3,v) - 
         4 i_2({\cal T}_4,v)\Big] \Big\} \nnb \\
\ar {e^{-m_b^2/M^2} \over 2304 \sqrt{3} \pi^2 m_b M^2}
\Big\{2 (2 - \beta - \beta^2) \GG m_b^2 (e_s \sp - e_u \uu) i_2({\cal S},v) \nnb \\
\ar (1 - \beta) \Big[144 (1 + \beta) f_{3\gamma} m_0^2 m_b^2 \pi^2 (e_u \sp - e_s \uu) 
       \widetilde{j}_1(\psi^v) \nnb \\
\ek 2 (3 + \beta) \GG m_b^2 (e_s \sp - e_u \uu) 
      \widetilde{j}_2(h_\gamma) + (e_u \sp - e_s \uu) \Big( (2 + \beta) \GG m_0^2 \nnb \\
\ek 2 (1 + 2 \beta) \GG m_b^2 - 36 (1 + \beta) f_{3\gamma} m_0^2 m_b^2 \pi^2 
        \psi^a(u_0)\Big)\Big]\Big\} \nnb \\
%
%
\ar {e^{-m_b^2/M^2} \over 1152 \sqrt{3} \pi^2 m_b}
(1 - \beta) (e_s \sp - e_u \uu) \Big[(1 - \beta) \GG + 
   18 (7 - \beta) m_b^2 e^{m_b^2/M^2} ({\cal I}_0 - m_b^2 {\cal I}_1) \nnb \\
\ek 18 (5 + \beta) m_b^2 e^{m_b^2/M^2} {\cal I}_{\ell n}\Big]
\Big[i_2({\cal S},1) - i_2({\cal T}_4,1)\Big] \nnb \\
\ar {e^{-m_b^2/M^2} \over 1152 \sqrt{3} \pi^2 m_b}
(1 - \beta) (e_s \sp - e_u \uu) \Big[(1 - \beta) \GG + 
    18 (1 - 7 \beta) m_b^2 e^{m_b^2/M^2} ({\cal I}_0 - m_b^2 {\cal I}_1) \nnb \\
\ar 18 (1 + 5 \beta) m_b^2 e^{m_b^2/M^2} {\cal I}_{\ell n}\Big] \Big[i_2(\widetilde{\cal S},1)
+ i_2({\cal T}_2,1)\Big] \nnb \\
\ar {e^{-m_b^2/M^2} \over 768 \sqrt{3} \pi^2 m_b}
(1 - \beta^2) (e_s \sp - e_u \uu) 
  \Big\{ \Big[\GG + 72 m_b^2 e^{m_b^2/M^2} ({\cal I}_0 - m_b^2 {\cal I}_1) \nnb \\
\ek 36 m_b^2 e^{m_b^2/M^2} {\cal I}_{\ell n}\Big] 
     i_2({\cal T}_1,1)) - \Big[\GG + 36 m_b^2 e^{m_b^2/M^2} {\cal I}_{\ell n}\Big] i_2({\cal
T}_3,1)\Big\} \nnb \\
\ek {e^{-m_b^2/M^2} \over 1152 \sqrt{3} \pi^2 m_b} 
(1 - \beta) (e_s \sp - e_u \uu) \Big[\GG + 36 m_b^2 e^{m_b^2/M^2} ({\cal I}_0 -
m_b^2 {\cal I}_1)\Big] \nnb \\
\cp  \Big\{2 (2 + \beta) i_2({\cal S},v) + i_2(\widetilde{\cal S},v) + 3 i_2({\cal T}_2,v) - 
   3 i_2({\cal T}_3,v) - \beta \Big[i_2(\widetilde{\cal S},v) - i_2({\cal T}_2,v) \nnb \\
\ar 3 i_2({\cal T}_3,v) - 
     2 i_2({\cal T}_4,v) \Big]\Big\} \nnb \\
%
%
\ek {e^{-m_b^2/M^2} \over 4608 \sqrt{3} \pi^2}
(e_s - e_u) f_{3\gamma} \Big\{(1 - \beta)^2 \GG - 
    72 e^{m_b^2/M^2} \Big[2 (1 + \beta + \beta^2) {\cal I}_{-1} - (1 - \beta)^2 m_b^2 {\cal I}_0 \nnb \\
\ek (1 + 4 \beta + \beta^2) m_b^4 {\cal I}_1\Big] - 216 (1 + \beta)^2 m_b^2
e^{m_b^2/M^2}  {\cal I}_{\ell n}\Big\} i_3({\cal V},v) \nnb \\
%
%
\ar {e^{-m_b^2/M^2} \over 4608 \sqrt{3} \pi^2}
(e_s - e_u) f_{3\gamma} \Big\{(1 - \beta)^2 \GG + 
   72 e^{m_b^2/M^2} \Big[(1 + 4 \beta + \beta^2) {\cal I}_{-1} + (1 - \beta)^2 m_b^2 {\cal I}_0 \nnb \\
\ek 2 (1 + \beta + \beta^2) m_b^4 {\cal I}_1\Big] + 216 (1 + \beta)^2 m_b^2 e^{m_b^2/M^2} 
    {\cal I}_{\ell n}\Big\} i_3({\cal A},v) \nnb \\
%
\ek {1\over 768 \sqrt{3} \pi^4 m_b^2}
(1 + \beta + \beta^2) (e_s - e_u) \Big\{12 {\cal I}_{-3} - 
   m_b^2 \Big[48 {\cal I}_{-2} - 72 m_b^2 {\cal I}_{-1} + (\GG + 48 m_b^4)
{\cal I}_0 \Big]\Big\} \nnb \\
%
%
\ek {1 \over 32 \sqrt{3} \pi^2}
(1 - \beta) m_b \Big[(7 + 2 \beta) e_b (\sp - \uu) - 
    \beta (3 e_u \sp - 3 e_s \uu)\Big] {\cal I}_0 \nnb \\
\ar {1 \over 128 \pi^2}
\sqrt{3} (1 - \beta^2) m_b ((e_s \sp) - e_u \uu) \mathbb{A} (u_0) {\cal I}_0 \nnb \\
%
%
\ar {1 \over 768 \sqrt{3} \pi^4}
m_b \Big\{-(1 + \beta + \beta^2) (e_s - e_u) m_b (\GG + 12 m_b^4) \nnb \\
\ek 3 (1 - \beta) e_b [(5 + \beta) m_0^2 - 24 (3 + \beta) m_b^2] \pi^2 (\sp - \uu) \nnb \\
\ar 48 (1 - \beta)^2 m_b^2 \pi^2 (e_u \sp - e_s \uu) - 
   18 (1 - \beta^2) m_b^2 \pi^2 (e_s \sp - e_u \uu) \mathbb{A} (u_0)\Big\} {\cal I}_1 \nnb \\
%
%
\ek {1 \over 384 \sqrt{3} \pi^2}
(1 - \beta) m_b 
  \Big\{ \Big[(2 + \beta) e_u \GG - 3 m_0^2 m_b^2 (3 + 2 \beta) e_b -
3 m_0^2 m_b^2 (7 + 5 \beta) e_u \nnb \\
\ar 12 (2 + \beta) (e_b + e_u) m_b^4\Big] \sp - 
   \Big[(2 + \beta) e_s \GG - 3 m_0^2  m_b^2 (3 + 2 \beta) e_b - 3 m_0^2
m_b^2 (7 + 5 \beta) e_s  \nnb \\
\ar 12 (2 + \beta) (e_b + e_s) m_b^4\Big] \uu \Big\} {\cal I}_2 \nnb \\
%
%
\ar {1 \over 384 \sqrt{3} \pi^2}
(1 - \beta) (5 + \beta) e_b m_b (\sp - \uu) {\cal I}_{\ell n} \nnb \\
%
%
\ar {e^{-m_b^2/M^2} \over 4608 \sqrt{3} \pi^4 m_b^2}
\Big\{ 4 (1 - \beta) m_b \pi^2 (e_u \sp - e_s \uu) \Big[(4 + 5 \beta) \GG -
18 (1 + 2 \beta) m_0^2 m_b^2\Big] \nnb \\
\ar 3 (1 - \beta^2) \GG m_b \pi^2 (e_s \sp - e_u \uu) 
   \mathbb{A} (u_0) \nnb \\
\ar 36 e^{m_b^2/M^2} (1 + \beta + \beta^2) (e_s - e_u) 
   \Big[2 {\cal I}_{-3} - 9 m_b^2 {\cal I}_{-2} + 18 m_b^4 {\cal I}_{-1} - 11 m_b^6 {\cal I}_0 + 
    6 m_b^6 {\cal I}_{\ell n}\Big] \Big\} \nnb \\
%
%
%
\ar {e^{-m_b^2/M^2} \over 6 \sqrt{3}}
f_{3\gamma} m_b (2 - \beta - \beta^2) (e_u \sp - e_s \uu)
\widetilde{j}_1(\psi^v) \nnb \\
%
%
\ar {e^{-m_b^2/M^2} \over 576 \sqrt{3} \pi^2 m_b}
(1 - \beta) (3 + \beta) (e_s \sp - e_u \uu) 
   \Big[\GG + 36 e^{m_b^2/M^2} m_b^2 ({\cal I}_0 - m_b^2 {\cal I}_1)
\Big] \widetilde{j}_2(h_\gamma) \nnb \\
%
%
\ek {1\over 768 \sqrt{3} \pi^2 m_b}
(1 - \beta^2) (e_s \sp - e_u \uu) \chi \Big\{2 \GG {\cal I}_0 + 
   3 m_b^2 \Big[12 {\cal I}_{-1} - (\GG + 12 m_b^4) {\cal I}_1\Big]\nnb \\
\ar 72 m_b^4 {\cal I}_{\ell n}\Big\} 
  \varphi_\gamma(u_0) \nnb \\
%
%
\ek {e^{-m_b^2/M^2} \over 24 \sqrt{3}}
f_{3\gamma} m_b (1 + \beta - 2\beta^2) (e_u \sp - e_s \uu)
\psi^a(u_0) \nnb \\
%
\ar {e^{-m_b^2/M^2}\over 1152 \sqrt{3} \pi^4} 
(1 + \beta + \beta^2) (e_s - e_u) f_{3\gamma} 
 \Big\{\GG - 36 e^{m_b^2/M^2} \Big[{\cal I}_{-1} - 2 m_b^2 {\cal I}_0 +
m_b^4 {\cal I}_1\Big] \Big\} \nnb \\
\cp  \Big[4 \widetilde{j}_1(\psi^v) - \psi^a(u_0)\Big]~. \nnb \\ \nnb \\
\eea
{\bf 4) Coefficient of the $(p\varepsilon) I$ structure}
\bea
&& \Pi_4^B = \nnb \\             
%
%
%
%
\ar {e^{-m_b^2/M^2}\over 3456 \sqrt{3} M^8}
(1 - \beta^2) f_{3\gamma} \GG m_0^2 m_b^4 (e_u \sp - e_s \uu) \psi^v(u_0) \nnb \\
%
%
\ek {e^{-m_b^2/M^2}\over 1728 \sqrt{3} M^6} 
(1 - \beta^2) f_{3\gamma} \GG m_0^2 m_b^2 (e_u \sp - e_s \uu) \psi^v(u_0) \nnb \\
%
%
\ar {e^{-m_b^2/M^2}\over 6912 \sqrt{3} \pi^2 M^4} 
(1 - \beta) \GG m_b^2 (e_u \sp - e_s \uu) \Big[3 (2 + \beta) m_0^2 - 
   8 (1 + \beta) f_{3\gamma} \pi^2 \psi^v(u_0)\Big] \nnb \\
%
%
\ar {e^{-m_b^2/M^2}\over 48 \sqrt{3} M^2}
(1 - \beta^2) f_{3\gamma} m_0^2 m_b^2 (e_u \sp - e_s \uu) \psi^v(u_0) \nnb \\
%
%
%
\ek {1 \over 64 \sqrt{3} \pi^2 m_b^2}
(1 - \beta) \Big[ (1 + \beta) e_b (\sp - \uu) + 
   4 (2 + \beta) (e_u \sp - e_s \uu) \nnb \\
\ek 12 (e_s \sp -
e_u \uu) \widetilde{j}_1(h_\gamma) \Big] {\cal I}_{-2} \nnb \\
%
%
\ar {1 \over 16 \sqrt{3} \pi^2}
(1 - \beta) m_b^2  \Big[(7 + 3 \beta) e_b (\sp - \uu) - 
   (3 + \beta) (e_s - e_u) f_{3\gamma} m_b \psi^v(u_0)\Big]
{\cal I}_{\ell n}\nnb \\
%
%
\ar {1 \over 32 \sqrt{3} \pi^2}
(1 - \beta)  \Big[(5 + 3 \beta) e_b (\sp - \uu) + 
   6 (2 + \beta) (e_u \sp - e_s \uu) \nnb \\
\ek 18 (e_s \sp - e_u \uu) \widetilde{j}_1(h_\gamma) -
   (3 + \beta) (e_s - e_u) f_{3\gamma} m_b \psi^v(u_0)\Big] {\cal I}_{-1} \nnb \\
%
%
\ar {1 \over 384 \sqrt{3} \pi^2}
(1 - \beta) m_b
  \Big\{6 m_b \Big[\beta e_u m_0^2 + 4 (2 + \beta) (e_b + e_u) m_b^2\Big] \sp \nnb\\
\ek 6 m_b \Big[\beta e_s m_0^2 + 4 (2 + \beta) (e_b + e_s) m_b^2\Big] \uu - 
   72 m_b^3 (e_s \sp - e_u \uu) \widetilde{j}_1(h_\gamma) \nnb \\
\ar (3 + \beta) (e_s - e_u) 
    f_{3\gamma} (\GG + 12 m_b^4) \psi^v(u_0)\Big\}  {\cal I}_1 \nnb \\
%
%
\ek {1 \over 576 \sqrt{3} \pi^2 m_b}
(1 - \beta) \Big\{ 9 m_b \Big[e_b \Big((3 + 2 \beta) m_0^2 + (17 + 9 \beta) m_b^2\Big) 
      (\sp - \uu) \nnb \\
\ar \Big((4 + 3 \beta) m_0^2 + 12 (2 + \beta) m_b^2\Big) 
      (e_u \sp - e_s \uu) - 36 m_b^2 (e_s \sp - e_u \uu) \widetilde{j}_1(h_\gamma)\Big] \nnb \\
\ar f_{3\gamma} \Big[(3 + \beta) (e_s - e_u) \GG - 
     48 (1 + \beta) e_u m_b \pi^2 \sp + 48 (1 + \beta) e_s m_b \pi^2 \uu\Big]
    \psi^v(u_0)\Big\}  {\cal I}_0 \nnb \\
%
%
\ar {e^{-m_b^2/M^2} \over 576 \sqrt{3} \pi^2 m_b^2}
(1 + \beta) \Big\{9 e^{m_b^2/M^2}  
    \Big[(1 + \beta) e_b (\sp - \uu) + 4 (2 + \beta) (e_u \sp - e_s \uu) \nnb \\
\ek 12 (e_s \sp - e_u \uu) \widetilde{j}_1(h_\gamma)\Big] 
({\cal I}_{-2} - 2 m_b^2 {\cal I}_{-1} + m_b^4 {\cal I}_0)
+ m_b^2 \Big[3 \GG (e_s \sp - e_u \uu) \widetilde{j}_1(h_\gamma) \nnb \\
\ek (e_u \sp - e_s \uu) \Big((2 + \beta) \GG - 2 (1 + \beta) f_{3\gamma} m_0^2 \pi^2 
        \psi^v(u_0)\Big) \Big] \Big\}~. \nnb
\eea

The functions $i_n~(n=1,2)$, and $\widetilde{j}_1(f(u))$
are defined as:
\bea
\label{nolabel}
i_0(\phi,f(v)) \es \int {\cal D}\alpha_i \int_0^1 dv
\phi(\alpha_{\bar{q}},\alpha_q,\alpha_g) f(v) (k-u_0) \theta(k-u_0)~, \nnb \\
i_1(\phi,f(v)) \es \int {\cal D}\alpha_i \int_0^1 dv
\phi(\alpha_{\bar{q}},\alpha_q,\alpha_g) f(v) \theta(k-u_0)~, \nnb \\
i_2(\phi,f(v)) \es \int {\cal D}\alpha_i \int_0^1 dv
\phi(\alpha_{\bar{q}},\alpha_q,\alpha_g) f(v) \delta(k-u_0)~, \nnb \\
i_3(\phi,f(v)) \es \int {\cal D}\alpha_i \int_0^1 dv
\phi(\alpha_{\bar{q}},\alpha_q,\alpha_g) f(v) \delta^\prime(k-u_0)~, \nnb \\
i_4(\phi,f(v)) \es \int {\cal D}\alpha_i \int_0^1 dv
\phi(\alpha_{\bar{q}},\alpha_q,\alpha_g) f(v) \delta^{\prime\prime}(k-u_0)~, \nnb \\
\widetilde{j}_1(f(u)) \es \int_{u_0}^1 du f(u)~, \nnb \\
\widetilde{j}_2(f(u)) \es \int_{u_0}^1 du (u-u_0) f(u)~, \nnb \\
{\cal I}_n \es \int_{m_b^2}^{\infty} ds\, {e^{-s/M^2} \over s^n}~,\nnb \\
{\cal I}_{\ell n} \es \int_{m_b^2}^{\infty} ds\, e^{-s/M^2} {\ell
n}{m_b^2\over s}~,\nnb
\eea
where 
\bea
k = \alpha_q + \alpha_g \bar{v}~,~~~~~u_0={M_1^2 \over M_1^2
+M_2^2}~,~~~~~M^2={M_1^2 M_2^2 \over M_1^2 +M_2^2}~.\nnb
\eea

\newpage

\section*{Figure captions}
{\bf Fig. (1)} The dependence of the  of the form factor $f_2^-$ on
$\cos\theta$ at six fixed values of the Borel mass parameter $M^2$,
and at the fixed value $s_0=42~GeV^2$ of the continuum threshold
for the $\Sigma_b^0 \to \Lambda_b \gamma$ transition.\\\\
{\bf Fig. (2)} The same as Fig. (1), but at the fixed value $s_0=44~GeV^2$
of the continuum threshold.

\newpage

\begin{figure}
\vskip 3. cm
    \includegraphics{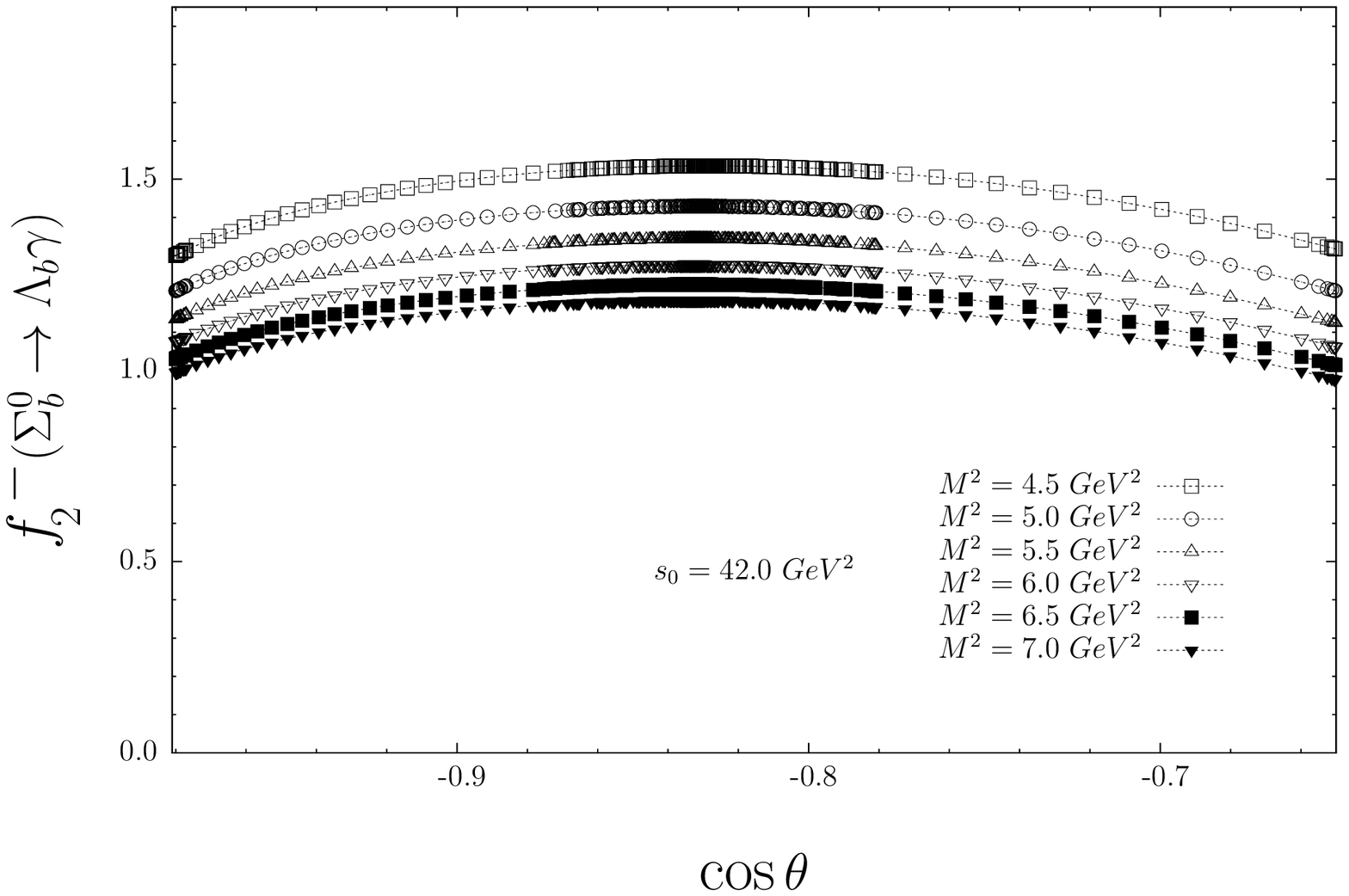}
\vskip 7.0cm
\caption{}
\end{figure}

\begin{figure}
\vskip 3. cm
    \includegraphics{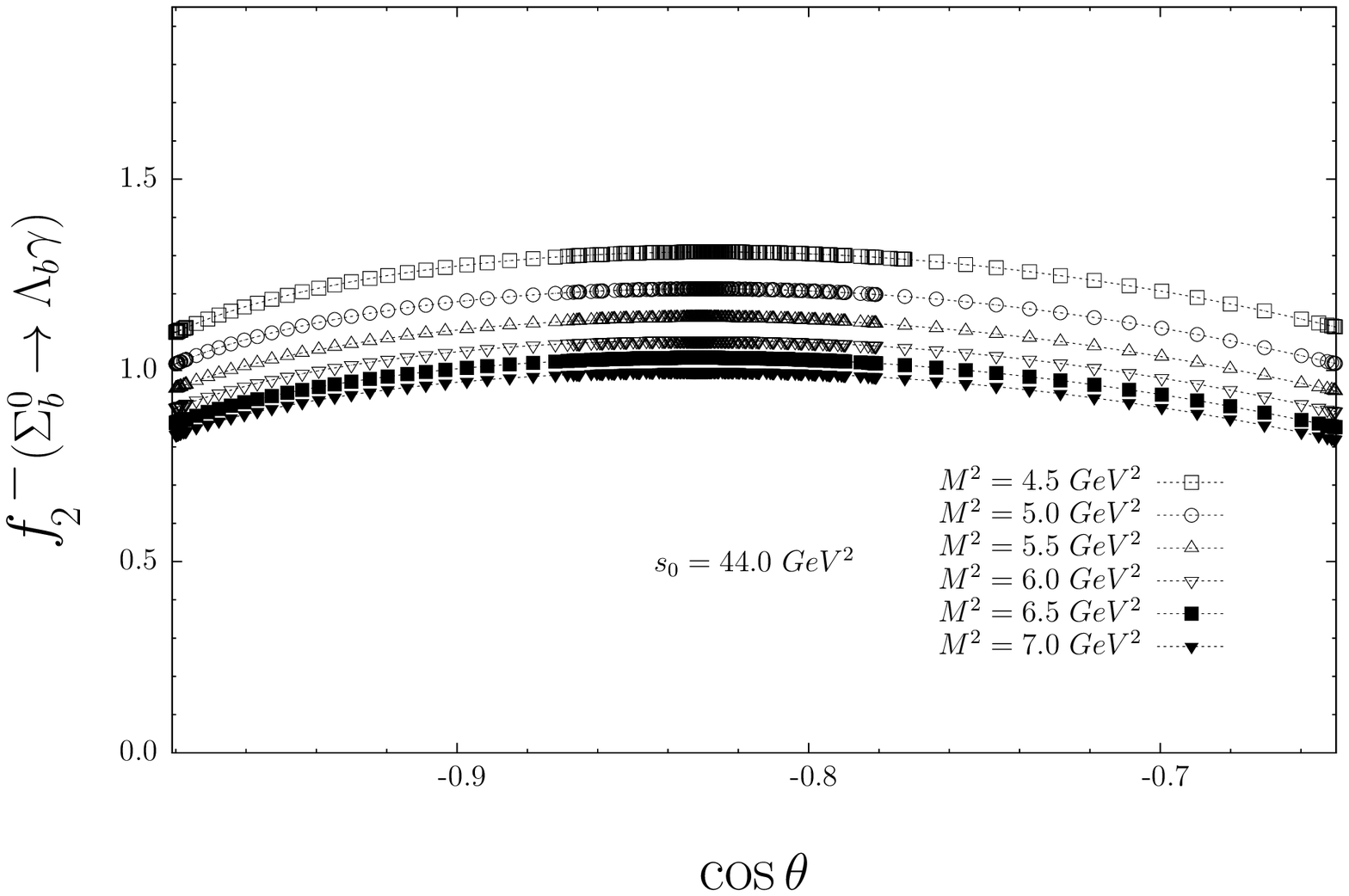}
\vskip 7.0cm
\caption{}
\end{figure}

\end{document}